\documentclass[12pt]{article}
\input epsf
\makeatletter

\def\newpic#1{}
\def\hybrid{\topmargin 0pt      \oddsidemargin 0pt
            \headheight 0pt \headsep 0pt

            \textwidth 6.25in       % A4 paper
            \textheight 9.5in       % A4 paper
            \marginparwidth 0.0in
            \parskip 5pt plus 1pt   \jot = 1.5ex}
\catcode`\@=11
\def\marginnote#1{}

\newcount\hour
\newcount\minute
\newtoks\amorpm
\hour=\time\divide\hour by60
\minute=\time{\multiply\hour by60 \global\advance\minute by-\hour}
\edef\standardtime{{\ifnum\hour<12 \global\amorpm={am}%
            \else\global\amorpm={pm}\advance\hour by-12 \fi
            \ifnum\hour=0 \hour=12 \fi
            \number\hour:\ifnum\minute<10 0\fi\number\minute\the\amorpm}}
\edef\militarytime{\number\hour:\ifnum\minute<10 0\fi\number\minute}

\def\draftlabel#1{{\@bsphack\if@filesw {\let\thepage\relax
       \xdef\@gtempa{\write\@auxout{\string
          \newlabel{#1}{{\@currentlabel}{\thepage}}}}}\@gtempa
       \if@nobreak \ifvmode\nobreak\fi\fi\fi\@esphack}
            \gdef\@eqnlabel{#1}}
\def\@eqnlabel{}
\def\@vacuum{}
\def\draftmarginnote#1{\marginpar{\raggedright\scriptsize\tt#1}}

\def\draftlabel#1{{\@bsphack\if@filesw {\let\thepage\relax
       \xdef\@gtempa{\write\@auxout{\string
          \newlabel{#1}{{\@currentlabel}{\thepage}}}}}\@gtempa
       \if@nobreak \ifvmode\nobreak\fi\fi\fi\@esphack}
            \gdef\@eqnlabel{#1}}
\def\@eqnlabel{}
\def\@vacuum{}
\def\draftmarginnote#1{\marginpar{\raggedright\scriptsize\tt#1}}

\def\draft{\oddsidemargin -.5truein
            \def\@oddfoot{\sl preliminary draft \hfil
            \rm\thepage\hfil\sl\today\quad\militarytime}
            \let\@evenfoot\@oddfoot \overfullrule 3pt
            \let\label=\draftlabel
            \let\marginnote=\draftmarginnote
       \def\@eqnnum{(\theequation)\rlap{\kern\marginparsep\tt\@eqnlabel}%
\global\let\@eqnlabel\@vacuum}  }

\def\numberbysection{\@addtoreset{equation}{section}
            \def\theequation{\thesection.\arabic{equation}}}

\def\underline#1{\relax\ifmmode\@@underline#1\else
            $\@@underline{\hbox{#1}}$\relax\fi}

\def\titlepage{\@restonecolfalse\if@twocolumn\@restonecoltrue\onecolumn
         \else \newpage \fi \thispagestyle{empty}\c@page\z@
            \def\thefootnote{\fnsymbol{footnote}} }

\def\endtitlepage{\if@restonecol\twocolumn \else  \fi
            \def\thefootnote{\arabic{footnote}}
            \setcounter{footnote}{0}}  %\c@footnote\z@ }
\relax

\makeatletter
\newdimen\normalarrayskip              % skip between lines
\newdimen\minarrayskip                 % minimal skip between lines
\normalarrayskip\baselineskip
\minarrayskip\jot
\newif\ifold             \oldtrue            \def\new{\oldfalse}
\def\arraymode{\ifold\relax\else\displaystyle\fi} % mode of array entries
\def\eqnumphantom{\phantom{(\theequation)}}     % right phantom in eqnarray
\def\@arrayskip{\ifold\baselineskip\z@\lineskip\z@
        \else
        \baselineskip\minarrayskip\lineskip2\minarrayskip\fi}
\def\@arrayclassz{\ifcase \@lastchclass \@acolampacol \or
\@ampacol \or \or \or \@addamp \or
      \@acolampacol \or \@firstampfalse \@acol \fi
\edef\@preamble{\@preamble
     \ifcase \@chnum
        \hfil$\relax\arraymode\@sharp$\hfil
        \or $\relax\arraymode\@sharp$\hfil
        \or \hfil$\relax\arraymode\@sharp$\fi}}
\def\@array[#1]#2{\setbox\@arstrutbox=\hbox{\vrule
        height\arraystretch \ht\strutbox
        depth\arraystretch \dp\strutbox
        width\z@}\@mkpream{#2}\edef\@preamble{\halign
\noexpand\@halignto
\bgroup \tabskip\z@ \@arstrut \@preamble \tabskip\z@ \cr}%
\let\@startpbox\@@startpbox \let\@endpbox\@@endpbox
     \if #1t\vtop \else \if#1b\vbox \else \vcenter \fi\fi
     \bgroup \let\par\relax
     \let\@sharp##\let\protect\relax
     \@arrayskip\@preamble}
\def\eqnarray{\stepcounter{equation}%
                 \let\@currentlabel=\theequation
                 \global\@eqnswtrue
                 \global\@eqcnt\z@
                 \tabskip\@centering
                 \let\\=\@eqncr
    \halign to \displaywidth\bgroup
       \eqnumphantom\@eqnsel\hskip\@centering
       $\displaystyle \tabskip\z@ {##}$%
       \global\@eqcnt\@ne \hskip 2\arraycolsep
            $\displaystyle\arraymode{##}$\hfil
       \global\@eqcnt\tw@ \hskip 2\arraycolsep
            $\displaystyle\tabskip\z@{##}$\hfil
            \tabskip\@centering
       &{##}\tabskip\z@\cr}
\begingroup\ifx\undefined\newsymbol \else\def\input#1 {\endgroup}\fi
\newfont{\hr}{msbm10}
\newfont{\ams}{msam10}
\def\beq{\begin{equation}}
\def\eeq{\end{equation}}
\def\ba{\beq\new\begin{array}{c}}
\def\ea{\end{array}\eeq}
\def\be{\ba}
\def\ee{\ea}

\numberbysection
\hybrid

\def\beq{\begin{equation}}
\def\eeq{\end{equation}}
\def\p{\partial}

\begin{document}
\begin{titlepage}

\title{Integrable Structure of the Dirichlet Boundary Problem
in Two Dimensions}

\author{A.Marshakov \thanks{Theory Department,
Lebedev Physics Institute, Leninsky pr.
53, 117924 Moscow, Russia and ITEP,
Bol. Cheremushkinskaya str. 25, 117259 Moscow, Russia}
\and P.Wiegmann \thanks{James Franck Institute and Enrico Fermi
Institute
of the University of Chicago, 5640 S.Ellis Avenue,
Chicago, IL 60637, USA and
Landau Institute for Theoretical Physics, Moscow, Russia}
\and A.Zabrodin
\thanks{Institute of Biochemical Physics,
Kosygina str. 4, 119991 Moscow, Russia
and ITEP, Bol. Cheremushkinskaya str. 25, 117259 Moscow, Russia}}

\date{September 2001}
\maketitle
\vspace{-7cm}

\centerline{
\hfill ITEP/TH-32/01}
\centerline{
\hfill FIAN/TD-11/01}

\vspace{7cm}

\begin{abstract}

We study how the solution of the two-dimensional
Dirichlet boundary problem for smooth simply connected
domains depends upon variations
of the data of the problem. We show that the Hadamard formula for the
variation of the Dirichlet Green function
under deformations of the domain
reveals an integrable structure.
The independent variables corresponding to the infinite set of
commuting flows are identified with harmonic moments of the domain. The
solution to the Dirichlet boundary problem is expressed through the
tau-function of the dispersionless Toda
hierarchy. We also discuss a degenerate case of the Dirichlet problem on
the plane with a gap. In this case the tau-function is identical
to the  partition function
of the planar large $N$ limit of the Hermitean one-matrix model.
\end{abstract}

\vfill

\end{titlepage}
\section{Introduction}

The subject of the Dirichlet boundary problem in two dimensions \cite{C-H}
is a harmonic function in a  domain
of the complex plane bounded by a closed
curve with a given value on the boundary and
continuous up to the boundary. The question we
address in this paper is how the harmonic function
in the bulk varies under a small
deformation of the
the shape of the domain.

Remarkably, this standard problem of complex
analysis possesses an integrable
structure \cite{M-W-Z,W-Z}
which we intend to clarify further in this paper.
It is described by a particular solution
of an integrable hierarchy of partial differential
equations known in the literature as dispersionless
Toda (dToda) hierarchy.
Moreover,
related integrable hierarchies
arise in the context of 2D topological
theories and just the same solution to the dToda hierarchy
emerges in the study of 2D quantum
gravity \cite{gravity,2matrix}
(we do not elaborate these relations in this paper).

Let $D$ be a simply connected domain in the complex plane
bounded by a smooth simple curve $\gamma$. The Dirichlet
problem is to find a harmonic
function $u(z)$ in $D$ such that it is continuous
up to the boundary and equals a given function
$u_0(z)$ on the boundary. The problem has
a unique solution written in terms of the
Green function $G(z_1,z_2)$
of the Dirichlet boundary problem:
\beq\label{G3}
u(z)=-\,
\frac{1}{2\pi}\oint_{\gamma}
u_0(\xi )\p_{n} G(z,\xi ) |d\xi |\,,
\eeq
where $\p_n$ is the normal derivative on the boundary
with respect to the second variable,
and the normal vector $\vec n$ always looks {\it inside} the domain,
where the Dirichlet problem is posed. Equivalently, the
solution is represented as
$u(z)=\displaystyle{\frac{1}{\pi i}\oint_{\p D}
u_0(\xi )\p_{\xi } G(z,\xi ) d\xi}$, where
$\p D$ is understood as $\gamma$ run
anticlockwise with respect to the domain.

The main object to study is, therefore, the Dirichlet Green function.
It is uniquely determined by the following properties \cite{C-H}:
\begin{itemize}
\item[(G1)] The function $G(z_1,z_2)-\log|z_1-z_2|$ is symmetric,
bounded and harmonic everywhere in $D$ in both arguments;
\item[(G2)] $G(z_1,z_2)=0$ if any
one of the variables belongs to the boundary.
\end{itemize}
The definition implies that $G(z_1, z_2)$ is real
and negative in $D$.
The Green function can be written explicitly through
a conformal map
of the domain $D$ onto some ``reference'' domain for which
the Green function is known. A convenient choice is the
unit disk. Let $f(z)$ be any bijective
conformal map of $D$ onto the unit disk (or its complement), then
\beq\label{G2}
G(z_1, z_2)=\log \left |
\frac{f(z_1)-f(z_2)}{f(z_1)\overline{f(z_2)} -1} \right |,
\eeq
where bar means complex conjugation. Such a map exists
by virtue of the Riemann mapping theorem \cite{C-H}.

It thus suffices to
study variations of the
conformal map $f(z)$ under deformations of the
boundary.
This problem was discussed in \cite{M-W-Z,W-Z}, where it
was shown that evolution of the conformal map
under changing harmonic moments of the
domain is given by the dToda
integrable hierarchy\footnote{A relation
between conformal maps (of slit domains)
and special solutions to some integrable
equations of hydrodynamic type was earlier observed
by Gibbons and Tsarev \cite{GT1}.}.
The study of the Dirichlet problem   approaches this subject
from another angle.

Our starting point
is the Hadamard variational formula \cite{Hadamard}. It
gives variation of the Green function under small deformations
of the domain in terms of the Green function itself:
\beq\label{Hadam}
\delta G(z_1, z_2)=\frac{1}{2\pi}\oint_{\gamma}
\p_{n}G(z_1, \xi)\p_{n}G(z_2, \xi)\delta h(\xi)|d\xi |.
\eeq
Here $\delta h(\xi)$ is the thickness between the curve
$\gamma$ and the deformed curve, counted along the normal
vector at the point $\xi \in \gamma$.
We show that already this remarkable formula reflects all integrable
properties of the Dirichlet problem.

A smooth closed curve  $\gamma$ (for simplicity, we
may assume it to be analytic in order to
have an easy sufficient justification of some arguments below)
divides the complex plane into two parts having the common boundary:
a compact interior domain $D_{\rm int}$,
and an exterior domain $D_{\rm ext}$ containing $\infty$.
Correspondingly, one
recognizes {\it interior} and {\it exterior}  Dirichlet
problems. The main contents of the paper is common
for both of them. To stress this, we try to keep
the notation uniform
calling the domain simply $D$.
We will show that (logarithms of) the tau-functions,  introduced
in \cite{M-W-Z,W-Z}
and further studied
in \cite{K-K-MW-W-Z},
for the interior and exterior problems are
related to each other by a Legendre transform.

The exterior Dirichlet problem makes sense when
the interior domain degenerates into a segment
(a plane with a gap). We will show that in this case a deformation
problem is described by the dispersionless limit of the Toda chain
hierarchy and discuss its relation
to the planar limit of the Hermitean matrix model.

\section{Deformations of the boundary
\label{ss:deform}}

Let $D$ be a simply-connected domain in the
extended complex plane bounded by a smooth simple
curve $\gamma$. Consider a basis $\psi_k(z)$,
$k\geq 1$, of holomorphic functions in $D$ such that
$\psi_k(z_0)=0$ for some  point $z_0 \in D$.
We call $z_0$ {\it the normalization point}.
The basis is assumed to be fixed and independent of the domain.
For example, in case of the interior problem one may
assume, without loss of generality, that the origin is in $D$
and set
$z_0 =0$, $\psi_k (z)=z^k/k$, while a natural choice
for the exterior problem is $z_0 =\infty$ and
$\psi_k(z)=z^{-k}/k$.
Throughout the paper, these bases for
interior and exterior problems are refered to as natural ones.

Let $t_k$ be moments of the domain $D$ defined with respect
to the basis $\psi_k$:
\beq\label{M4}
t_k= \, \kappa\frac{1}{\pi }
\int_{D}\psi_k (z) \,d^2 z\,, \,\;\;\;\;\;k=1,2,\ldots
\eeq
where $\kappa=\pm$  for the
interior (exterior) problem. We also assume
that the functions $\psi_k$ for domains containing $\infty$
are integrable, or the integrals are properly regularized
(see below). Besides, we denote by $t_0$
the area (divided by $\pi$) of the domain $D$
in case of the interior problem and that of the
complementary (compact) domain in case of the exterior
problem:
$$
t_0 =\left \{
\begin{array}{ll}
\displaystyle{\frac{1}{\pi}\int_{D}d^2z} &
\mbox{for compact domains}
\\&\\
\displaystyle{\frac{1}{\pi}\int_{{\bf C}\setminus D}d^2z} &
\mbox{for non-compact domains}\end{array}\right.
$$
Let us note that the
moments (except for $t_0$) are in general complex.
We call the quantities $t_k$, $\bar t_k$ and $t_0$
{\it harmonic moments} of the domain $D$.
The Stokes formula
represents the harmonic moments as contour
integrals
$$
t_{k}=\frac{1}{2\pi i }\oint_{\gamma}
\psi_k(z)\bar z dz\,,
\;\;\;\;\;\;k=0,1,2,\ldots
$$
(where it is set $\psi_0(z)=1$)
providing, in particular, a regularization of possibly divergent
integrals (\ref{M4}) in case of the exterior problem.
Throughout the paper
the contour in $\oint_{\gamma}$ is
run in anticlockwise direction both for interior and
exterior problems.

The basic fact of the theory of deformations of
closed analytic curves is that the (in general
complex) moments $t_k$
supplemented by the real variable
$t_0$ form a set of local coordinates in
the space of smooth closed curves \cite{Kriunp} (see also
\cite{T}).
This means that under any small deformation of the domain the set
$\{t_0,t_1,\ldots\}$ is subject to a small change and vice
versa.
More precisely, let $\gamma(t)$ be
a family of curves such that
$\partial_t t_k=0$ in some neighborhood of
$t=0$, then all the curves $\gamma (t)$ coincide with
$\gamma=\gamma(0)$ in this neighborhood.

The family of differential operators
\beq\label{D2}
\nabla (z)= \p_{t_0}
+\sum_{k\geq 1}\Bigl (\psi_k (z)\p_{t_k} +
\overline{\psi_k (z)}\p_{\bar t_k}\Bigr )
\eeq
span the complexified tangent space to the space of
curves.
They are invariant under change of variables in the
following sense:
let $\tilde t_k$ be harmonic moments
defined with respect to another basis, $\tilde \psi_k$,
of holomorphic functions in $D$; then
$\tilde \nabla (z)=\nabla (z)$. Note that
$\nabla (z_0)=\p_{t_0}$ since $\psi_k(z_0)=0$.
The operator $\nabla (z)$
has a clear geometrical meaning described below.

\begin{figure}[tp]
\epsfysize=4cm
\centerline{\epsfbox{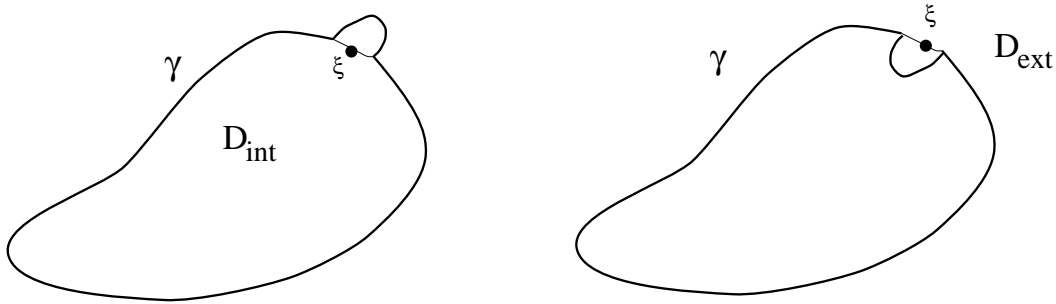}}
\caption{\sl Action of the operator $\nabla (\xi)$
in the case of interior $D_{\rm int}$ (left) and
exterior $D_{\rm ext}$ (right) domains.
In our convention bump always looks outside the (interior or
exterior) domain.}
\label{fi:bump}
\end{figure}

Let us consider a special deformation of the
domain obtained by adding to it an infinitesimal
smooth bump (of an arbitrary
form) with area $\epsilon$ located at the point
$\xi \in \gamma$.
Our convention is that $\epsilon >0$ if the bump
looks outside the domain in which the Dirichlet problem
is posed, as is shown in Fig.~\ref{fi:bump}.

Let $A$ be any functional of a domain that depends
on the harmonic moments only. The variation of such a functional
   in the leading order in $\epsilon$,
is given by
\beq\label{D4}
\delta_{\epsilon(\xi)} A =\kappa \frac{\epsilon }{\pi} \nabla
(\xi )A\,, \;\;\;\;\; \xi \in \gamma,
\eeq
Indeed, combining\quad
$\delta A = \p_{t_0}A \delta t_0 +
\displaystyle{\sum_{k\geq 1}( \p_{t_k}A \delta t_k +
\p_{\bar t_k}A
\delta \bar t_k )}$ and
$$\displaystyle{\delta t_k =\kappa
\frac{1}{\pi}\int_{\rm bump} \psi_{k}(z)d^2z =
\kappa\frac{\epsilon }{\pi }\psi_{k}(\xi)}$$
we obtain (\ref{D4}).
So, the result of the action of the operator $\nabla (\xi )$
with $\xi \in \gamma$ on $A$
is proportional to the variation of the
functional under attaching a bump at the point $\xi$.
To put it differently, we can say that the
boundary value of the function
$\nabla (z)A$ is
given by the l.h.s. of (\ref{D4}).
For functionals $A$ such that the series
$\nabla (z)A$ converges everywhere in $D$ up to the
boundary,
this remark gives a usable method to find the function
$\nabla (z)A$ everywhere in the domain.
This function is
harmonic in $D$ with the boundary value determined
from (\ref{D4}). It is given by (\ref{G3}):
\beq\label{DDD}
\frac{\epsilon}{\pi}\nabla (z)A=
\kappa \frac{1}{2\pi}\oint_{\gamma}|d\xi |
\p_{n} G(z,\xi ) \delta_{\epsilon(\xi)} A.
\eeq
This gives the result of the action of the operator
$\nabla (z)$,
when the argument is anywhere in $D$.

For example,
given any regular function $f$ in a domain
containing the interior domain $D$, set $A_f =\int_{D}f(z)d^2z$.
We have: $\displaystyle{
\delta_{\epsilon(\xi)}A_f
=\frac{\epsilon}{\pi}\nabla (\xi)
A_f= \epsilon f(\xi )}$,
$\xi \in \gamma$.
If the function $f$ is
harmonic in $D$, then
$\nabla (z )A_f =\pi f(z )$
for any $z \in D$.

The subject of the deformation
theory of the Dirichlet boundary problem is to
compute $\nabla (z)G(z_1,z_2)$
through the conformal map or the Green function
of the original domain. In the next section,
we do this using the Hadamard variational formula.

\section{Hadamard variational formula and
dispersionless integrable hierarchy}

\subsection{The
Hadamard integrability condition}

Variation of the Green function under
small deformations of the domain is
known due to Hadamard \cite{Hadamard}, see eq.\,(\ref{Hadam}).
Being specified
to the particular case of attaching a bump of
the area $\epsilon$, it reads:
\beq\label{D8}
\delta_{\epsilon(\xi)}G(z_1,z_2 )=
-\, \frac{\epsilon}{2\pi} \p_{n}G(z_1,\xi )
\p_{n}G(z_2 ,\xi ),\;\;\;\;\;
\xi\in\gamma.
\eeq

To find how the Green function changes under a
variation of the harmonic moments, we use (\ref{DDD})
to employ the harmonic continuation procedure
explained in the previous section.
The harmonic function in $D$
with a boundary value given by the Hadamard formula is
\beq\label{Th0}
\nabla (z_3)G(z_1, z_2) =
\kappa\frac{1}{4\pi }\oint_{\gamma}
\p_n G(z_1 , \xi) \p_n G(z_2 , \xi)
\p_{n} G(z_3 , \xi) |d\xi|
\eeq

It is obvious from the r.h.s. of (\ref{Th0}) that the
result of the
action of the operator $\nabla (z)$ on the Green function
is harmonic and symmetric in all three arguments, i.e.,
\beq\label{1}
\nabla (z_3)G(z_1,z_2)
=\nabla (z_1)G(z_2,z_3).
\eeq
This is our basic relation.
It has the form of integrability condition.
In the rest of the paper we will draw
consequences of this symmetry and
underlying algebraic structures.
Note also that despite the Green function vanishes on the boundary, its
derivative (the
l.h.s. of  eq.\,(\ref{1})) with respect to the deformation of the domain
does not.

The basic equation
(\ref{1}) is a compressed
form of an integrable hierarchy.
To unfold it,
let us separate holomorphic and
antiholomorphic parts of this equation.

Let $E$ be the exterior to the unit disk.
Given a point
$a\in D$, consider a bijective
conformal map $f_{a}:D\rightarrow E$ such that
$f_a(a )=\infty$.
The Dirichlet Green function then is
\be
\label{Gfplus}
G(a,z)=-\log |f_{a}(z)|.
\ee
Under a proper normalization of  the map
the integrability condition (\ref{1}) becomes holomorphic:
\beq\label{hie2}
\nabla (b) \log f_{a}(z)
=\nabla (a) \log f_{b}(z)
\eeq
for all $a,b,z\in D$.

The following normalization will be  convenient: the overall phase
is chozen to be
$
\mbox{arg} f_a\,(z_0) =\pi -\mbox{arg}\,(z_0 -a)$ if $a\neq z_0
$,
where $z_0\in D$ is the normalization point. If $a=z_0$ we set
$\displaystyle{\lim_{z\to z_0}\Bigl
((z-z_0)^2f'(z) \Bigr )}$ to be real and negative.
Under these conditions
\beq\label{hie1}
f_{a}(z)= \left ( \frac{(\bar a-\bar z_0)\,\,
f(a)}{(a - z_0)\,\,\overline{f(a)}}\right )^{1/2} \!
\frac{f(z)\overline{f (a)} -1}{f(a)-f(z)}
\eeq
(for $a, z_0 \neq \infty$).
In the vicinity of  the point $a$
($a\neq \infty$)
\beq\label{hie00}
f_a(z)=e^{i\omega (a,z_0)}\,\frac{r_a}{z\!-\!a}\,\Bigl (
1+\sum_{k\geq 1}p_k(a)(z\!-\!a)^k \Bigr ),
\eeq
where the real constant $r_a$ is called conformal radius
of the domain \cite{Hille}
with respect to the point $a$, and $\omega$
is a phase determined from the normalization condition.
In particular, they read $\omega (z_0 , z_0)=0$.
Similarly, the map $f_a(z)$ can be defined
in the case when either $a$ or $z_0$ lies
at infinity.

To verify (\ref{hie2}), we note that the holomorphic function
$\nabla (b) \log f_{a}(z)
-\nabla (a) \log f_{b}(z)$ is also antiholomorphic
(in $z$) by virtue
of (\ref{1}), and thus must be a constant.
Setting  $z=z_0$ we find that the latter is zero:
$$
\begin{array}{lll}
&&
\nabla (b) \log |f_a(z_0)|
-\nabla (a) \log |f_b(z_0)| \,+
\\&&\\
&+&
i\nabla (b) \,\mbox{arg}\, f_a(z_0)
-i\nabla (a) \,\mbox{arg}\, f_b(z_0)\, =0
\end{array}
$$
(the first line vanishes due to (\ref{1}), the second one vanishes
because  the normalization does not depend on the
shape of the domain).

\subsection{Harmonic moments as commuting flows}

Equation (\ref{hie2}) suggests to treat
$\log f_{a}(z)$ as a generating function
of commuting flows with respect to spectral parameter $a$.
The expansion
\beq\label{F1}
\log f_{a}(z)=
H_{0}(z)+\sum_{k\geq 1}\Bigl (\psi_k(a) \,H_k(z)-
\overline{\psi_k (a)} \,\tilde H_k(z)\Bigr )
\eeq
defines generators
$H_{k}$, $\tilde H_k$
of the commuting flows.
Clearly, $H_{0}(z)=\log f(z)$.
It implies evolution
equations for $f(z)$,
$$
\frac{\p \log f(z)}{\p t_k}=
\frac{\p H_{k}(z)}{\p t_0}\,,
\;\;\;\;\;\;
\frac{\p \log f(z)}{\p \bar t_k}=-
\frac{\p \tilde H_{k}(z)}{\p t_0}
$$
and integrability conditions:
\beq\label{F888}
\frac{\p H_{k}(z)}{\p t_j}=
\frac{\p H_{j}(z)}{\p t_k}\,,
\;\;\;\;\;
\frac{\p \tilde H_{k}(z)}{\p \bar t_j}=
\frac{\p \tilde H_{j}(z)}{\p \bar t_k}\,,
\;\;\;\;\;
\frac{\p H_{k}(z)}{\p \bar t_j}=
-\, \frac{\p \tilde H_{j}(z)}{\p t_k}\,.
\eeq
The real part of (\ref{F1}) vanishes on the boundary
     (as it is the Dirichlet Green function),
therefore, the boundary
values of  $\tilde H_{k}$ and $H_{k}$
are complex conjugated:
\beq\label{F88}
\tilde H_{k}(z)=\overline{H_{k}(z)}\,,
\;\;\;\;\;\; z\in \gamma.
\eeq

The structure of integrable hierarchy becomes explicit
if instead of functions of $z$
one passes to functions of its
image $w$ under the map $f_{z_0}$:
$w=f_{z_0}(z) \equiv f(z)$.

Using the chain rule, one can write
$$
\left. \phantom{a^A}
\nabla (a) \log f_b(z)=
\nabla (a)\log f_b(z(w))\right |_{w}
+\left(\nabla (a)\log f(z)\right)
w\p_{w}\log f_b.
$$
In the last term we observe that
$\nabla (a) \log f(z)=\p_{t_0}
\log f_{a}(z)$ (using (\ref{hie2}) at $b=z_0$).
Subtracting the same equality with $a,b$ interchanged,
we come to equation of the zero-curvature type:
$$
\nabla (a ) \log f_{b}
-\nabla (b ) \log f_{a} - \{
\log f_a ,\, \log f_b\} = 0,
$$
where the Poisson brackets are defined as
$\displaystyle{ \{f,g\}\equiv
w\frac{\p f}{\p  w}
\frac{\p g}{\p t_0}
-w\frac{\p g}{\p  w}
\frac{\p f}{\p t_0}}$ and $t_k$-derivatives are taken now at
fixed $w$.

Let $z(w)$ be the map inverse to $w=f(z)$.
Equation (\ref{hie2}) at $b=z_0$, being rewritten
for the inverse map, has the form of a
one-parametric family of evolution equations
of the Lax type labeled by the spectral
parameter $a$.
They are
\beq\label{F3}
\nabla (a) z(w)
=\{\log f_{a}(z(w)) ,\, z(w)\}.
\eeq
We refer to them as to {\it deformation equations}.
The zero-curvature conditions ensure that these equations
are consistent, i.e. flows with
different values of spectral parameter commute.

The integrability conditions (\ref{F888}) in the new
variable acquire the form of the
zero-curvature
equations
\beq\label{65}
\begin{array}{l}
\p _{t_j} H_i(w)
-\p _{t_i}H_j(w) +\{H_i(w),\,H_j(w)\}=0,
\\ \\
\p _{\bar t_j} \tilde H_i(w)
-\p _{\bar t_i}\tilde H_j(w) +\{\tilde H_j(w),\,\tilde H_i(w)\}=0,
\\ \\
\p _{t_j}\tilde H_i(w)
+\p _{\bar t_i}H_j(w) +\{\tilde H_i(w),\,H_j(w)\}=0.
\end{array}
\eeq
  From (\ref{hie1}) it
is easy to see that $H_k$ are polynomials in $w$
while $\tilde H_k$ are polynomials
in $w^{-1}$. Furthermore, for any basis such that
$\psi_k(z)= O(\,(z\!-\!z_0)^k \,)$ these polynomials are of
degree $k$.
In other words, the generators are
meromorphic functions
on the Riemann sphere with two marked points at $w=0$
and $w=\infty$. This is a particular case of the universal
Whitham hierarchy \cite{KriW}, known as dispersionless Toda lattice.
In \cite{Takashi}, it is
proved that the full set of zero-curvature conditions
(\ref{65}), together with the polynomial structure of
the generators, already imply existence of the Lax function
and Lax-Sato equations.

For a particular choice of the basis, the Lax function
can be expressed through the inverse conformal map.
Consider the interior problem, set $z_0 =0\in D$ and fix
the basis of holomorphic functions in $D$ to be the natural one:
$\psi_k(z)=z^k/k$.
 From (\ref{hie00}) and (\ref{F1}) it follows that
$\tilde H_k$ are holomorphic in $D$ while
$H_k$ are meromorphic with the $k$-th order
pole at $0$ so that
$H_{k}=z^{-k}+O(1)$ as $z\to 0$.
Combining these properties with the polynomial structure
of the generators as functions of $w$, and taking into
account (\ref{F88}), one gets
\beq\label{H}
\begin{array}{l}
H_k=\displaystyle{\Bigl (z^{-k}(w)\Bigr )_{>0} +\frac{1}{2}
\Bigl (z^{-k}(w) \Bigr )_{0}},
\\ \\
\displaystyle{\tilde H_k=\Bigl (\bar z^{-k}(w^{-1})\Bigr )_{<0}
+\frac{1}{2}
\Bigl (\bar z^{-k}(w^{-1}) \Bigr )_{0}},
\end{array}
\eeq
where $z(w)$ is the map inverse to $f(z)$
and $\bar z(w)=\overline{z(\bar w})$ is its
Schwarz double.
The symbols $(f(w))_{>0}$, $(f(w))_{<0}$
mean truncated Laurent series, where
only terms with strictly positive (negative)
powers of $w$ are kept, while $(f(w))_{0}$ is the
free term ($w^0$) of the series.
The free terms in (\ref{H}) are found
with the help of an easily proved identity
$\displaystyle{\log \frac{w(z)}{z}=\sum_{k\geq 1}\frac{1}{k}
(z^{-k}(w))_{0}z^k}$ valid for any series of the form
$w(z)=z+w_1 z^2 +\ldots$ .
Similar formulas hold for the exterior problem with the
choice $z_0 =\infty$, $\psi_k (z)=z^{-k}/k$.

\subsection{Deformation equations and dispersionless Toda
hierarchy}

Expending the deformation equations (\ref{F3}) in  spectral parameter we
obtain
the Lax representation of the dToda hierarchy.
To see this, we recall that the dToda hierarchy is an infinite
set of evolution (Lax-Sato) equations for two Lax functions
$$
\begin{array}{l}
\displaystyle{L(w)=rw \,+\,\sum_{k\geq 0}u_k w^{-k}},
\\ \\
\displaystyle{\tilde L(w)=rw^{-1}+
\sum_{k\geq 0}\tilde u_k w^{k}}.
\end{array}
$$
The equations are
$$
\frac{\p L(w)}{\p t_k}=\{ H_k (w), \, L(w)\}\,,
\;\;\;\;\;\;
\frac{\p L(w)}{\p \bar t_k}=\{ L(w), \, \tilde H_k(w)\}
$$
and the same for $\tilde L(w)$.
Here the Poisson brackets with respect to $w$ and
$t_0$ are defined as above, and
$$
\begin{array}{l}
\displaystyle{
H_k (w)=\Bigl (L^k(w)\Bigr )_{>0}+
\frac{1}{2} \Bigl (L^k(w)\Bigr )_{0}}
\\ \\
\displaystyle{
\tilde H_k (w)=\Bigl (\tilde L^k(w)\Bigr )_{<0}+
\frac{1}{2} \Bigl (\tilde L^k(w)\Bigr )_{0}}
\end{array}
$$
are generators of the flows. One may also introduce
$H_0(w) =\log w$.
They obey
the dispersionless zero-curvature
conditions (\ref{65})
which express consistency of the Lax equations and
generate an infinite set of partial differential equations
for coefficients of the Lax functions.

To summarize, we have the following
identification of the Lax functions
with conformal maps:
\begin{itemize}
\item For the interior problem
($z_0 =0$ with the natural basis):
\beq\label{toda4}
L(w)=\frac{1}{z(w)}\,,
\;\;\;\;\;
\tilde L(w)=\frac{1}{\bar z(w^{-1})}\,,
\eeq
where $z(w)$ is the function inverse to
$w=f_{0}(z)$;
\item For the exterior problem
($z_0 = \infty$ with the natural basis):
\beq\label{toda5}
L(w)=z(w)\,,
\;\;\;\;\;
\tilde L(w)=\bar z(w^{-1})\,,
\eeq
where $z(w)$ is the function inverse to
$w=f_{\infty}(z)$
\end{itemize}
So the inverse map, $z(w)$, and its Schwarz double,
$\bar z(w^{-1})$, both obey the Lax equations.

Let us comment on another choice of basis.
Suppose $\lambda^{-1}$ is a local parameter at the normalization
point $z_0$, i.e.,
$$
\begin{array}{ll}
\displaystyle{
\lambda = \frac{1}{z\!-\!z_0}+c_0 +c_1 (z\!-\!z_0)+
O(\,(z\!-\!z_0)^2)}\,,
& z_0 \neq \infty
\\&\\
\displaystyle{
\lambda =z+c_0 + \frac{c_1}{z}+O(z^{-2})}\,,
& z_0 =\infty.
\end{array}
$$
with some domain-independent coefficients $c_j$.
Assuming that $\lambda^{-1} (z)$ is a well defined local
parameter in a domain that contains $D$, we choose
the basis of holomorphic functions in $D$ to be
$\psi_k(z)=\lambda^{-k}(z)/k$. This results in
a linear change of times with the help of a triangular
matrix.
Repeating the above arguments, one identifies the Lax
function with $\lambda(z)$: $L(w)=\lambda (z(w))$.
This Lax function provides a solution to an equivalent
hierarchy in the sense of \cite{Shiota}.

\subsection{String equations}

A customary way to fix a solution to a dispersionless
hierarchy is to impose
an additional constraint on the Lax functions (sometimes called
{\it string equation} \cite{Douglas}). We are going to show that
not only the deformation equations but the string
equation, too, is an easy consequence of the
Hadamard formula and our basic relation (\ref{1}).

By $r=r_{z_0}$ denote the conformal radius of the
curve $\gamma$ with respect to the normalization point $z_0$.
Let us calculate $\delta_{\epsilon (\xi)} \log r$ in two
different ways. The first one is to use the Hadamard formula
(\ref{D8}) when both arguments tend to the normalization
point:
$$
\delta_{\epsilon (\xi)} \log r =
\frac{2\epsilon}{\pi} | \p_{\xi} G(z_0, \xi)|^2
=\frac{\epsilon}{2\pi} | \p_{\xi} f(\xi)|^2
$$
(recall that $|f(\xi)|=1$ for $\xi \in \gamma$).
The second one is obtained from (\ref{1}) in the limit
$z_1 \to \xi \in \gamma$, $z_2 , z_3 \to z_0$:
$$
\delta_{\epsilon (\xi)} \log r
=-\,\frac{\epsilon}{\pi}
\p_{t_0}\log \left | f(\xi)\right |,
$$
where we have used (\ref{D4}).
Combining the results, we obtain the relation
$$
\p_{t_0}\log  | f(z) |^2 =
- | \p_{z} f(z) |^2\,,
\;\;\;\;\;\;z\in \gamma.
$$

Passing to the variable $w=f(z)$, one
rewrites this equation as
$$
2{\cal R}e \left (w\, \frac{\p z(w)}{\p w}\,
\frac{\p \overline{z(w)}}{\p t_0}\right ) =1\,,
\;\;\;\;\;\; |w|=1\,,
$$
where $z(w)$ is the inverse map, as before.
Being analytically continued from the unit circle,
it reads
$$
\{z(w),\, \bar z(w^{-1})\}=1.
$$
This is the customary form of the semiclassical
string equation.
Similar arguments in case of the exterior problem lead to
the same equation.

For the natural basis, let us rewrite the string
equation
in terms of the Lax functions $L, \tilde L$.
Using the identifications (\ref{toda4}) and (\ref{toda5}),
we have:
\beq\label{string3}
\begin{array}{ll}
\{L^{-1}(w),\, \tilde L^{-1}(w)\}=-1
&\;\; \mbox{for the interior problem}
\\ & \\
\{L(w),\, \tilde L(w)\}=1
&\;\; \mbox{for the exterior problem}
\end{array}
\eeq
So, although the string equation in terms of $z(w)$
is the same,
the interior and exterior problems correspond to
two different solutions of the same (dToda) integrable hierarchy
(cf.\,\cite{Crete}).

\section{Tau-function and dispersionless Hirota equations}

\subsection{Tau-function}

Symmetry relation (\ref{1})
implies that there exists a
real-valued function of harmonic moments
$F(t_0, {\bf t}, \bar {\bf t})=
F(t_0; t_1,t_2,\ldots ; \bar t_1 , \bar t_2 ,\ldots)$
such that
$$
G(z_1,z_2 )=g_{0}(z_1, z_2) +
\frac{1}{2}\nabla (z_1)\nabla (z_2)F,
$$
where the function $g_0$ does not depend on the domain, i.e.,
on moments (the coefficient $\frac{1}{2}$ is set for
future convenience). Note that adding to $F$ a quadratic
form in $t_k$, $\bar t_k$ amounts to changing the function
$g_0$ only.

  From the definition of the Green function it follows that $g_0
(z_1, z_2)$ is symmetric and harmonic in $D$ in both arguments with
the only singularity $\log |z_1 -z_2|$ as $z_1 \to z_2$,
or, in terms of a local parameter $\lambda^{-1}$,
$\log |\lambda^{-1}(z_1)-\lambda^{-1}(z_2)|$.
In case of the interior problem,
$D$ can be any domain not containing the point of infinity. The function
$g_0$ must be the same for all such domains.
Therefore, singularities of $g_0$
in the extended complex plane, other than the logarithmic
singularity at merging points,
may occur only at infinity.
In other words, $g_0$ is of the form
$g_0 (z_1 , z_2) =
\log |\lambda^{-1}(z_1) -\lambda^{-1}(z_2)| + h_0(z_1, z_2)$
where $h_0$ is regular and harmonic in both arguments
everywhere in the complex plane but at infinity.
The function $h_0$
is in fact a matter of definition of the tau-function.
For some particular basis we are free to choose it to be zero.
Hence we obtain the following important relation:
\beq\label{sec3a}
G(z_1,z_2 )=\log |\lambda^{-1}(z_1) -\lambda^{-1}(z_2)| +
\frac{1}{2}\nabla
(z_1)\nabla (z_2 )F \,,
\;\;\;\;z_{1}, z_{2} \in D
\eeq
Changing the normalization point or passing to another local
parameter results in modifying the function $g_0$. As is
already pointed out, this is equivalent to adding to $F$
a quadratic form in times.
Taking into account the remark
at the end of Sec.\,3.4, one sees
that this agrees with the relation between tau-functions
of equivalent hierarchies discussed in \cite{KMMM}.
It is easy to see that in this form the relation holds true
for the exterior problem as well.

For clarity,
let us specify the above equation for the natural basis.
Instead of the differential
operator $\nabla (z)$ defined for arbitrary basis,
it is convenient to use its specialization for the natural basis
\beq\label{calD}
{\cal D}(z)=\p_{t_0}+\sum_{k\geq 1}
\left (\frac{z^{-k}}{k}\p_{t_k}
+\frac{\bar z^{-k}}{k}\p_{\bar t_k}\right )
\eeq
so that $\nabla (z)={\cal D}(z)$ for the exterior problem and
$\nabla (z)={\cal D}(z^{-1})$ for the interior one.
Eq.\,(\ref{sec3a})
is then rewritten in the form
\beq\label{sec3}
G(z_1,z_2 )=\left \{
\begin{array}{ll}
\displaystyle{
\log |z_1 - z_2| +
\frac{1}{2}{\cal D}(z_1^{-1})
{\cal D}(z_2^{-1})F}&\mbox{for the interior problem}
\\& \\
\displaystyle{
\log  |z_1^{-1} - z_2^{-1} | +
\frac{1}{2}{\cal D}(z_1) {\cal D}(z_2 )F}&
\mbox{for the exterior problem}
\end{array}
\right.
\eeq

The function $F$ plays a central role in what follows.
It is the tau-function of curves
discussed in \cite{M-W-Z,W-Z,K-K-MW-W-Z}.
It is a dispersionless limit of logarithm of tau-function
of an integrable hierarchy or {\it dispersionless tau-function}
\cite{KriW}. For  brevity we  refer to
it simply as tau-function.
A relation between the tau-function
and the Green function of the type (\ref{sec3}) has been
emphasized by L.Takhtajan \cite{T}.
In sect.~\ref{ss:integral}
we give a few explicit integral representations for the
tau-function.

\subsection{Conformal maps and tau-function}

In this section, we use the natural basis.
To unify formulas for the interior and exterior cases,
we write
$\psi_k(z)=\lambda^{-k}(z)/k$, where
$\lambda(z)=z^{-1}$ for the interior problem
and $\lambda(z)=z$ for the exterior one.
(Correspondingly, $z_0$ is either $0$ or $\infty$.)
We also employ the notation
$f(z_j)=f_j$, $\lambda (z_j)=\lambda_j$,
\beq\label{DD}
D(z)=\sum_{k\geq 1}\frac{z^{-k}}{k}
\frac{\p}{\p t_k}\,,
\;\;\;\;\;\;
\bar D(\bar z)=\sum_{k\geq 1}\frac{\bar z^{-k}}{k}
\frac{\p}{\p \bar t_k},
\eeq
so that ${\cal D}(z)=\p_{t_0}+D(z)+\bar D(\bar z)$.

Representations of the conformal map through the tau-function can be
obtained from (\ref{sec3}) by  separating holomorphic and antiholomorphic
parts in $z_1$ and $z_2$ of the Green function. Holomorphic parts in both
variables yield
\beq\label{sec5a}
\log\, \frac{f_{1}
-f_{2}}{\lambda_1 -\lambda_2}
=D(\lambda_1)D(\lambda_2)F -\frac{1}{2}\p_{t_0}^{2}F.
\eeq
Terms holomorphic in $z_1$ and antiholomorphic in $z_2$ yield
\beq\label{511a}
\log \left (1- \frac{1}{f_1\overline{f_2}}\right )=
-D(\lambda_1)\bar D(\overline{\lambda_2})F.
\eeq

Various specifications of (\ref{sec3}), (\ref{sec5a})
and (\ref{511a})
lead to representations of the conformal maps through the tau
function.
Tending, for example,
$z_2\to z_0$ in (\ref{sec3})
and extracting
holomorphic parts in $z_1$, we get the following
expressions for the conformal maps through the
tau-functions:
\beq\label{sec4a}
f(z)=\lambda(z)\, \exp \,\Bigl ( -\,\frac{1}{2}\p_{t_0}^{2}F
- \p_{t_0}D(\lambda(z))F \Bigr ).
\eeq
The leading coefficients as $z\to z_0$ give
formulas for the conformal radius:
\beq\label{sec7}
2\log r=-\kappa \p_{t_0}^{2}F
\eeq
(recall that the sign factor $\kappa$ is introduced
in Sec.\,2 to distinguish between interior and exterior
problems).
Limits $z_2 \to z_0$ in (\ref{sec5a}), (\ref{511a})
give other representations for the conformal maps,
\beq\label{1qa}
\begin{array}{l}
\displaystyle{f(z)=
e^{-\frac{1}{2}\p_{t_0}^{2}F}
\left (\lambda(z) -\p^2_{t_0 t_1}F-
D(\lambda(z))\p_{t_1} F\right )},
\\ \\
\displaystyle{\frac{1}{f(z)}=
e^{-\frac{1}{2}\p_{t_0}^{2}F}
D(\lambda(z))\p_{\bar t_1} F}
\end{array}
\eeq
and another representation of the conformal radius:
\beq\label{rr}
r^{-2\kappa}=\p^2_{t_1 \bar t_1}F.
\eeq
These formulas agree with (\ref{sec7}) provided
$F$ satisfies the dispersionless Toda equation
\beq
\label{Toda}
\p^2_{t_1 \bar t_1}F =e^{\p_{t_0}^{2}F}.
\eeq
In fact all coefficients of the Taylor expansion
of the conformal map around the normalization point
can be easily read from (\ref{1qa}).

Merging the points $z_1$ and $z_2$ in
(\ref{sec5a}), (\ref{511a}), we get formulas for
the derivative and the  modules of the conformal maps:
\beq\label{ll}
\begin{array}{l}
\displaystyle{
f'(z)=\lambda '(z)e^{D^2(\lambda(z))F-\frac{1}{2}\p_{t_0}^{2}F}}
\\ \\
\displaystyle{
1-|f(z)|^{-2}=\exp \left (-D(\lambda(z))
\bar D(\overline{\lambda(z)})F \right )\,.}
\end{array}
\eeq

It is easy to
see that the generators of commuting flows introduced in (\ref{F1})
are expressed through the tau-function as follows:
$$
\begin{array}{l}
H_{k}(z)=\displaystyle{\lambda^k(z)-\Bigl (
\frac{1}{2} \p_{t_0} +D(\lambda(z))\Bigr ) \p_{t_k}F},
\\ \\
\tilde H_{k}(z)=\displaystyle{\Bigl (
\frac{1}{2} \p_{t_0} +D(\lambda(z))\Bigr ) \p_{\bar t_k}F},
\\ \\
H_{0}(z)=\displaystyle{\log \lambda(z) -\Bigl (
\frac{1}{2} \p_{t_0} +D(\lambda(z))\Bigr ) \p_{t_0}F}.
\end{array}
$$
As is seen from (\ref{sec3}),
coefficients of the Laurent series of the regular part of the Green
function are second order derivatives of the tau-function.
The formula
$$
2(z_1\p_{z_1} +z_2 \p_{z_2} )G(z_1,z_2) =
\frac{1}{2\pi i}\oint_{\p D}
\frac{d\log f_{z_1}(z)\, d\log f_{z_2}(z)}{d\log z}
$$
can be easily proved by substituting (\ref{Gfplus})
and evaluating residues
at the poles $z_1, z_2$. The left hand side of this formula
could be thought of as a dispersionless
analog of the Gelfand-Dikii resolvent.
Expanding both sides in power series in $z_1 , z_2$
using (\ref{sec3}) and (\ref{F1}), we obtain
$$
\begin{array}{l}
\displaystyle{
\frac{\p^2 F}{\p t_n \p t_m}\,=\,\,
\frac{1}{2\pi i (n\!+\!m)} \oint_{\gamma}
\frac{dH_{n} \, dH_{m}}{d\log z}}\,,
\;\;\;\;\;\;n\geq 0,\;\; m\geq 1
\\ \\
\displaystyle{
\frac{\p^2 F}{\p t_n \p \bar t_m}=-\,
\frac{1}{2\pi i \,n}\, \oint_{\gamma}
\frac{dH_{n} \,d\overline{H_{m}} }{d\log z}}\,,
\;\;\;\;\;\;\;\;\;\;n,m\geq 1
\end{array}
$$
(in the last formula we used (\ref{F88})).

In addition to the relations above, we point out that
the tau-function provides new
representations for some classical objects of
complex analysis.
Consider, for example,
the {\it Schwarz derivative}
${\cal S}(f,z)=
\frac{f'''(z)}{f'(z)}-\frac{3}{2} \left (
\frac{f''(z)}{f'(z)}\right )^2$
of the conformal map $f$. A non-singular part of
$\displaystyle{\lim_{\zeta \to z}\p_z \p_{\zeta}G(z, \zeta)}$
is $\frac{1}{12}{\cal S}(f,z)$,
so equation (\ref{sec3}), with
$z_0=\infty$ and $\lambda(z)=z$, yields~\cite{W-Z}
$$
{\cal S}(f, z)=6(\p_z D(z))^2 F.
$$
where $\p_z D(z)=-\displaystyle{\sum_{k\geq 1}
z^{-k-1}\p_{t_k}}$.
The reproducing kernel \cite{Hille} of the (exterior) domain
$D$,
$K(z, \bar \zeta ) =
2\p_z \p_{\bar \zeta}G(z, \zeta)$,
is given by
\beq\label{reprod}
K(z, \bar \zeta ) =\p_z D(z)
\p_{\bar \zeta}\bar D(\bar \zeta )F.
\eeq

\subsection{Dispersionless Hirota equations}

The tau-function obeys an infinite set of non-linear
differential equations.
They can be obtained by
excluding $f$ from the identities of the type
(\ref{sec5a}, \ref{511a}) by means of
eq.\,(\ref{sec4a}).
Equations obtained in this way are dispersionless
limits of the Hirota bilinear
relations \cite{Sato} obeyed by the tau-functions of
integrable hierarchies.

For example, consider eq.\,(\ref{sec5a})
for pairs of points $(z_1,z_2)$,
$(z_2 , z_3)$ and $(z_3 , z_1)$.
Exponentiating and summing the equations
one gets the identity
\beq\label{Hir1}
(z_1-z_2 )e^{D(z_1)D(z_2)F}
+(z_2-z_3 )e^{D(z_2)D(z_3)F}
+(z_3-z_1 )e^{D(z_3)D(z_1)F}=0.
\eeq
obeyed by
the tau-function $F$ for both interior and exterior
problems.
This is the symmetric form of
Hirota's equation for the dKP
hierarchy. It encodes algebraic relations
between the second order derivatives of the tau-function.
These relations are obtained on expanding the Hirota
equation in powers of $z_1, z_2, z_3$ and comparing
coefficients\footnote{In \cite{C-K,T-T},
the Hirota equation for the dKP
hierarchy was obtained in an equivalent but less
symmetric form. It follows from (\ref{Hir1}) in the
limit $z_3 \to \infty$.}
The operators $D(z)$ and
$\bar D(\bar z)$ are defined in (\ref{DD}).
Equation (\ref{Hir1}), as well as other Hirota equations
given below, are the same for interior and exterior problems.

More general equations  obtained in a similar
way include derivatives with respect to $t_0$
and $\bar t_k$. These are equations of the dToda
hierarchy:
\beq\label{Hir3} (z_1-z_2)
e^{D(z_1)D(z_2)F}
=z_1e^{-\p_{t_0}D(z_1)F}
-z_2 e^{-\p_{t_0}D(z_2)F},
\eeq
\beq\label{Hir4}
1-e^{-D(z_1)\bar D(\bar z_2 )F}=\frac{1}{z_1\bar z_2}
e^{\p_{t_0}(\p_{t_0}+ D(z_1)+\bar D(\bar z_2 ))F}.
\eeq
Note that eq.\,(\ref{Hir3}) can be obtained from (\ref{Hir1})
in the formal limit $z_3  \to 0$ if to understand
$\lim_{z \to 0}D(z)$ as $-\p_{t_0}$.
These equations allow one to express the second order
derivatives
$\p^2_{t_m t_n}F$,
$\p^2_{t_m \bar t_n}F$ with $m,n\geq 1$ as certain
functions of the derivatives
$\p^2_{t_0 t_k}F$,
$\p^2_{t_0 \bar t_k}F$.
The dispersionless Toda equation (\ref{Toda})
is the limit of (\ref{Hir4}) as $z_1, z_2 \to \infty$. Each side of this
equation is the squared conformal radius as one can see from
(\ref{sec7}, \ref{rr}).

In the same manner one can derive other equivalent forms
of the Hirota equations:
\be
\label{Hir5}
z_1(z_3 -z_2 )e^{-{\cal D}(z_3 )
(D(z_1)-D(z_3))F}-
z_2 (z_3 -z_1 )e^{-{\cal D}(z_3 )
(D(z_2)-D(z_3))F}\,=
\\
=\, z_3 (z_1 -z_2 )e^{(D(z_1)-D(z_3))(D(z_2)-D(z_3))F},
\ee
\be
\label{Hir6}
1-e^{-(D(z_1)-D(z_3))(\bar D(\bar z_2 )
-\bar D(\bar z_3 ))F} \,=
\\
=\, (z_{1}^{-1}-z_{3}^{-1})
(\bar z_{2}^{-1}-\bar z_{3}^{-1})
e^{{\cal D}(z_3 )({\cal D}(z_3 )+
D(z_1)-D(z_3 )+\bar D(\bar z_2 )-\bar D(\bar z_3 ))F},
\ee
where the operator ${\cal D}(z)$ is defined in (\ref{calD}).
Eqs.\,(\ref{Hir5}), (\ref{Hir6}) may be interpreted
as analogs of eqs.
(\ref{Hir3}, \ref{Hir4}) with the normalization point
moved to $z_3$, in which case $\p_{t_0}$ should be
substituted by ${\cal D}(z_3)$.
In the limit $z_3 \to \infty$
they convert into
eqs.\,(\ref{Hir3}), (\ref{Hir4}) respectively.
In (\ref{Hir6}) $\bar z_3$ can be regarded as an independent
formal variable, not necessarily complex conjugate to $z_3$.
We stress that the full set of differential equations for $F$
obtained by expanding  the Hirota equations
(\ref{Hir3}, \ref{Hir4}) or (\ref{Hir5},\ref{Hir6})
is the same.

The special case of (\ref{Hir6}) $z_2 =z_1$
is worth mentioning:
$$
1-e^{-(D(z)-D(\zeta ))(\bar D(\bar z)
-\bar D(\bar \zeta ))F} =
|z^{-1}-\zeta^{-1} |^2
e^{{\cal D}(\zeta ){\cal D}(z)F}.
$$
Further specialization $\zeta \to z$ yields
\beq\label{Hir8}
|z|^4 \p_z D(z)\p_{\bar z}\bar D (\bar z)F=
e^{{\cal D}^2(z)F},
\eeq
This equation is especially remarkable. On the one hand,
it looks like the dispersionless Toda
equation (\ref{Toda}) where
$\p_{t_0}$, $\p_{t_1}$, $\p_{\bar t_1}$ are replaced by
${\cal D}(z )$, $z^2 \p_z D(z)$,
$\bar z^2 \p_{\bar z}\bar D(\bar z)$ respectively
(moreover, (\ref{Hir8}) becomes (\ref{Toda}) as $z\to \infty$).
On the other hand, (\ref{Hir8}) is
equivalent to the Liouville equation. Indeed,
eq.\,(\ref{Hir8}) tells us that the field
$$
\chi (z)={\cal D}^2 (z)F -\log |z|^4
$$
obeys the Liouville equation
$$
\p_z \p_{\bar z}\chi =2e^{\chi}
$$
for $z\in D$. (Here we imply that $D$ contains $\infty$,
but similar equations can be written for the interior problem, too.)
By virtue of (\ref{sec3}),
(\ref{ll}), the solution to this equation can be
written as follows:
$$
\chi (z)=2\lim_{z'\to z}\left [
G(z,z')-\log |z-z'|\right ]
$$
or
$$
e^{\chi (z)}=\frac{|f'(z)|^2}{(|f(z)|^2 -1 )^2}
$$
Note that $ds^2 = e^{\chi}dz d\bar z$ is the
pull back of the Poincare metric
with constant negative curvature
$R_{ds^2}=-2e^{-\chi}\p_z \p_{\bar z}\chi =-4$ in $D$.
We also note that the Liouville field $\chi$
equals the value
of the reproducing kernel \cite{Hille}
of the domain $D$ at merging points:
$\chi (z)=\log K (z, \bar z)$, where $K(z, \bar \zeta)$
is the reproducing kernel (\ref{reprod}).

\subsection{Residue formulas}

In this section we present formulas for third
order derivatives of $F$.
Formulas of this type are known in the theory of dispersionless
integrable hierarchies and  are referred to as residue formulas
\cite{KriW}. They are used, in particular, to prove the associativity
equations for tau-functions of Whitham hierarchies
\cite{WDVV,MMM,BMRWZ}.

The basic relation to derive the residue formulas
is (\ref{Th0}).
By virtue of (\ref{sec3a}),
its left hand side is a generating function for
third order derivatives of the tau-function.
Let us first rewrite this formula in holomorphic terms.
To do that we note that $\p_n G(z, \xi)=-2|\p_{\xi}G(z,\xi)|$
for $\xi$ on the boundary (the vector
$\mbox{grad}\, G$ is normal to the boundary).
Further, since $G$ vanishes on the boundary,
$\p_{\xi} G (z, \xi)d\xi +\p_{\bar \xi} G (z, \xi)d\bar \xi =0$.
Therefore,
$|\p_{\xi}Gd\xi |=-\kappa i \p_{\xi}Gd\xi$ holds on the boundary.
Taking all this into account, we rewrite (\ref{Th0}) as
$$
\nabla (a)
\nabla (b)
\nabla (c) F
=\frac{4}{\pi i}\oint_{\gamma}
\frac{\p_z G(a,z)\p_z G(b,z)
\p_z G(c,z)}{dz d\bar z}\, (dz)^3.
$$
Finally, using $G(a,z)=-\log |f_{a}(z)|$,
where $f_{a}(z)$ is a holomorphic functions of $z$
(see (\ref{hie1})), we get
$$
\nabla (a)
\nabla (b)
\nabla (c) F
=\,
-\, \frac{1}{2\pi i}\oint_{\gamma}
\frac{d\log f_{a}(z)\, d\log f_{b}(z)
\,d\log f_{c}(z)}{dz d\bar z}.
$$
Let us expand the both sides in the basis $\psi_k$
in each argument
using (\ref{D2}), (\ref{F1}). Comparing the coefficients,
we come to
the residue formulas for third order derivatives
of the tau-function:
\beq\label{res4}
\begin{array}{l}
\displaystyle{
\frac{\p^3 F}{\p t_{l} \p t_m \p t_n}\,=\,
-\, \frac{1}{2\pi i} \oint_{\gamma}
\frac{dH_{l} dH_{m} \, dH_{n}}{dz d\bar z}}\,,
\;\;\;\;\;\;l,m,n \geq 0,
\\ \\
\displaystyle{
\frac{\p^3 F}{\p t_{l} \p t_m \p \bar t_n}=
\frac{1}{2\pi i }\, \oint_{\gamma}
\frac{dH_{l} dH_{m} \,
d\overline{H_{n} } }{dz d\bar z}}\,,
\;\;\;\;\;\;l,m\geq 0\,,\;\; n\geq 1.
\end{array}
\eeq
The formulas (\ref{res4}) were first obtained by I.Krichever \cite{Kriunp}
by other means.

\section{Integral representations of the tau-function
\label{ss:integral}}

In this section we discuss integral representations of the
tau-functions and connection between the tau-functions for the interior
and exterior Dirichlet problems.
We employ the special notation $D_{\rm int}$ for the interior
domain bounded by the curve $\gamma$ and $D_{\rm ext}$ for the
exterior one.
We use the natural basis, i.e.,
the normalization points are $0\in D_{\rm int}$
and $\infty \in D_{\rm ext}$, and the basis functions
are $z^k/k$ and $z^{-k}/k$ respectively.
At last, the tau-functions are denoted as $F^{\rm int}$,
$F^{\rm ext}$.

\subsection{Electrostatic potentials}

Let us begin with the exterior problem.
Our starting
point is eq.\,(\ref{sec3}) which we rewrite
here introducing the auxiliary function (potential)
\be
\label{phipm}
\Phi^{\rm ext}(z) = {\cal D}(z)F^{\rm ext}\,.
\ee
Eq.\,(\ref{sec3}) then reads
$$
{\cal D}(z)\Phi^{\rm ext}(\xi) = 2G(z,\xi )
- 2\log\left|z^{-1} - \xi^{-1}\right|
$$
for both $z$ and $\xi$ being in $D_{\rm ext}$.
(We recall that ${\cal D}(z)=\p_{t_0}+D(z)+\bar D(\bar z)$.)
This function admits an electrostatic interpretation
explained below.
When one of the points reaches the boundary,
the Green function vanishes and one has
$$
{\cal D}(\xi)
\Phi^{\rm ext} (z)=
-2 \log |z^{-1}-\xi^{-1}|,\;\;\;\;\; \xi \in \gamma.
$$
The last formula makes sense of the variation of
the function $\Phi^{\rm ext}$ under the bump deformation
of the domain (see sect.~\ref{ss:deform}).
According to
(\ref{D4}), the variation is
$$
\delta_{\epsilon (\xi )}\Phi^{\rm ext} (z)=
\frac{2\epsilon}{\pi}\log |z^{-1}-\xi ^{-1}|
$$
and, therefore,
one may write the integral representation
\beq\label{intre1}
\Phi^{\rm ext}(z) =
- {2\over\pi}\int_{D_{\rm int}}
d^2\zeta\log\left|z^{-1} - \zeta^{-1}\right|
\eeq
up to a harmonic function that do not depend on shape of the domain.
Adding to (\ref{intre1})
harmonic functions independent of times amounts to
redefinition of
the tau-function by terms linear in times, which
contribute neither to the
Green functions nor to the
formulas for conformal maps (\ref{sec4a}), (\ref{1qa}),
and leave Hirota equations unchanged.
We set all
such terms to vanish.
In a similar way, for the interior problem we get
$\delta_{\epsilon (\xi )}\Phi^{\rm int}(z)=
\displaystyle{\frac{2\epsilon}{\pi}}\log |z-\xi |$ and
$$
\Phi^{\rm int}(z) =
   {2\over\pi}\int_{D_{\rm ext}}d^2\zeta\log\left|z - \zeta\right| - c,
$$
where
the constant $c$ is necessary for regularization
of the divergent integral over $D_{\rm ext}$. Let us cut off the
integral at a big circle of radius $R$ and set
$c = c(R) = \displaystyle{
{2\over\pi}\int_{|\zeta|<R}d^2\zeta\log |\zeta |} =$
$R^2\log R^2 -R^2$.
Then  we get
\be
\label{intre2}
\Phi^{\rm int}(z) =
\lim_{R\to\infty}\left(
    {2\over\pi}\int_{D_{\rm ext}}
d^2\zeta\log\left|z - \zeta\right| - c(R)\right) =
{2\over\pi}\int_{D_{\rm ext}}d^2\zeta \log |1 - z\zeta^{-1}| -v_0,
\ee
where we introduced
$$
v_0=\frac{1}{\pi}\int_{D_{\rm int}}  \log|z|^2 \, d^2z.
$$

Formulas (\ref{intre1}), (\ref{intre2})
present the potentials
$\Phi^{\rm ext}$, $\Phi^{\rm int}$
in the form of an integral over the {\em complementary}
domain, since by definition (\ref{phipm})
the functions $\Phi^{\rm ext}$, $\Phi^{\rm int}$
are harmonic in $D_{\rm ext}$, $D_{\rm int}$ respectively.
One may use these formulas
to extend the functions $\Phi$
to the whole complex plane. For example, (\ref{intre1})
defines a function $\Phi^{\rm ext}$ which is
harmonic in the exterior domain and satisfies
the Poisson equation $-\p_z \p_{\bar z}
\Phi^{\rm ext}(z)=1-\pi t_0\delta(z)$
in $D_{\rm int}$. This function is  the potential
generated by a uniformly distributed charge in $D_{\rm int}$
and a compensating point-like charge at the origin. The expansions
of  $\Phi^{\rm ext}$ at small and large $z$
read:
\beq\label{expan12}
\begin{array}{ll}
\displaystyle{
\Phi^{\rm ext} (z)=-|z|^2+
2t_0\log |z|+\sum_{k>0}(t_{k}^{\rm ext} z^k
+\bar t_{k}^{\rm ext}\bar z^k )}\,,
&z\to 0,
\\ \\
\displaystyle{
\Phi^{\rm ext} (z)=v_0
+\sum_{k>0}(t_{k}^{\rm int} z^{-k}
+\bar t_{k}^{\rm int} \bar z^{-k})}\,,
&z\to \infty,
\end{array}
\eeq
where the additional superscripts are set to distinguish between
exterior and interior moments.
Under our assumptions, these series converge in $D_{\rm int}$
and $D_{\rm ext}$ respectively. The functions $\Phi^{\rm ext}$
and $\p_z \Phi^{\rm ext}$ are
continuous at the boundary.

\subsection{Integral formulas for the tau-function}

Using the same
strategy, we set $z$ in
(\ref{phipm}) to the boundary and interpret this formula
as a result of the bump deformation of the
domain. It is easy to check that the variation  of
\be
\label{first}
F^{\rm ext} = \frac{1}{2\pi}
\int_{D_{\rm int}} \!\!\!\! d^2 z \Phi^{\rm ext}(z)=
-\frac{1}{\pi^2}\int_{D_{\rm int}}d^2 z\int_{D_{\rm int}}d^2 \zeta\
\log \left |\frac{1}{z}-\frac{1}{\zeta}\right |
\ee
is
$$
\begin{array}{ll}
&\displaystyle{
\delta_{\epsilon (\xi )}F^{\rm ext} =
\frac{\epsilon}{2\pi} \Phi^{\rm ext} (\xi  )+
\frac{1}{2\pi}\int_{D_{\rm int}}d^2z
\delta_{\epsilon (\xi )}\Phi^{\rm ext} (z)\,=}
\\ & \\
=&
\displaystyle{
\frac{\epsilon}{2\pi} \Phi^{\rm ext} (\xi )-
\frac{\epsilon}{\pi^2}\int_{D_{\rm int}}d^2z \log |z^{-1} -\xi^{-1}|\,=
\frac{\epsilon}{\pi}  \Phi^{\rm ext} (\xi ).}
\end{array}
$$
Eq.\,(\ref{first}) presents the tau-function for the exterior
Dirichlet problem as a double integral over the
domain complement to the
$D_{\rm ext}$ \cite{K-K-MW-W-Z}.

Similar arguments give the tau-function of the interior problem:
$$
F^{\rm int} = \lim_{R\to\infty}
\left(-\,{1\over\pi^2}
\int_{D_{\rm ext}}d^2z\int_{D_{\rm ext}}
d^2\zeta\log\left|z-\zeta\right| +
C(R) - c(R)t_0\right)
$$
with $c(R)$ as in ({\ref{intre2}), and
$$
C(R) =
{1\over\pi^2}\int_{|z|<R}d^2z\int_{|\zeta| <R}d^2\zeta\log
\left|z-\zeta\right| =
{1\over 4}R^4(2\log R^2-1).
$$
It is implied that the integral over $D_{\rm ext}$ is cut off
at $R$.
Note the relation
\beq\label{55}
F^{\rm int} +F^{\rm ext}
=\frac{1}{\pi^2}\int_{D_{\rm int}}\!\!d^2z
\int_{D_{\rm ext}}\!\!d^2\zeta
\,\log |1-z\zeta^{-1}|^2
\eeq

In fact, the tau-functions admit other
useful integral representations.
Let us mention some of them.
In terms of the potential
$\Phi^{\rm ext}$, extended to the whole
complex plane as explained above, the tau-function of the exterior
problem may be represented as a (regularized) integral over
the whole complex plane:
\be
\label{second}
F^{\rm ext} = \lim_{\varepsilon \to 0}\left(
\frac{1}{2\pi}\int_{{\bf C}\setminus D_{\varepsilon}}
\left | \p_z \Phi^{\rm ext} \right |^2 d^2z +t_{0}^{2}\log\varepsilon
\right)
\ee
where $D_{\varepsilon}$ is a small disk of radius
$\varepsilon$ centered at the origin.
In the electrostatic interpretation, the tau-function
is basically the energy of the system of charges mentioned
above.
Stokes formula gives an integral representation  through
a contour integral:
\beq\label{299}
F^{\rm ext} = - \frac{1}{4}t_0^2+
\frac{1}{2\pi}\oint_{\gamma}
\Phi^{\rm ext}(z)\,\frac{\bar zdz-zd\bar z}{4i}
\eeq
Similar formulas can be written for the interior problem.

\subsection{Legendre transform $F^{\rm int}
\leftrightarrow F^{\rm ext}$
\label{ss:Legendre}}

The tau-functions for the interior and exterior problems
are connected by a Legendre transform.
The former is the function of the interior
moments $t_{k}^{\rm int}$ and the area $t_0$, while the latter is  the
function of the exterior
moments $t_{k}^{\rm ext}$ and the area.
From (\ref{phipm}) we read that first order derivatives of the
tau-function for the Dirichlet problem in $D_{\rm int}$ or
$D_{\rm ext}$
with respect to the harmonic moments are (up to the factor $k$) harmonic
moments of the complimentary domain:
\beq\label{first11}
\frac{\p F^{\rm int}}{\p t^{\rm int}_{k}}=
kt_{k}^{\rm ext}\,,
\;\;\;\;\;
\frac{\p F^{\rm ext}}{\p t^{\rm ext}_{k}}=
kt_{k}^{\rm int}
\eeq
and
$$
\frac{\p F^{\rm ext}}{\p t_{0}}=
-\frac{\p F^{\rm int}}{\p t_{0}}= v_0.
$$
Using (\ref{55}), we obtain the relation
$$
F^{\rm int}(t_0,{\bf t}^{\rm int}, \bar{\bf t}^{\rm int}) =
\sum_{k=1}^\infty
\left(k\,t^{\rm int}_k \,t^{\rm ext}_k +
k\,\bar t^{\rm int}_k \,\bar t^{\rm ext}_k\right)
-F^{\rm ext}(t_0,{\bf t}^{\rm ext}, \bar{\bf t}^{\rm ext}).
$$
By virtue of (\ref{first11}), it means that the functions
$F^{\rm int}$ and $F^{\rm ext}$ are related by
Legendre transforms
with respect to all variables but $t_0$:
\beq\label{Legendre}
\begin{array}{l}
\displaystyle{
F^{\rm int}=\sum_{k=1}^{\infty}\left (
t_{k}^{\rm ext}\, \frac{\p F^{\rm ext}}{\p t_{k}^{\rm ext}}
+\bar t_{k}^{\rm ext}\,
\frac{\p F^{\rm ext}}{\p \bar t_{k}^{\rm ext}} \right )
-F^{\rm ext} },
\\ \\
\displaystyle{
F^{\rm ext}=\sum_{k=1}^{\infty}\left (
t_{k}^{\rm int}\, \frac{\p F^{\rm int}}{\p t_{k}^{\rm int}}
+\bar t_{k}^{\rm int}\,
\frac{\p F^{\rm int}}{\p \bar t_{k}^{\rm int}} \right )
-F^{\rm int} }.
\end{array}
\eeq

\subsection{Homogeneity properties of the tau-function}

Expansion of the potential
$\Phi^{\rm ext}(z)$ around the origin (\ref{expan12})
allows one to prove another important property
of the tau-functions.
Integrating both sides of it over $D_{\rm int}$
and using (\ref{first11}), we obtain a quasihomogeneity
condition for $F^{\rm ext}$. A similar condition for $F^{\rm int}$
is most easily derived from (\ref{Legendre}). They are:
\beq\label{hom1}
\begin{array}{l}
\displaystyle{
4F^{\rm int} =
t_0^2 +2 t_0 \p_{t_0}F^{\rm int} +
\sum_{k>0}(2+k)(t_k \p_{t_k}F^{\rm int}
+\bar t_k \p_{\bar t_k}F^{\rm int} )},
\\ \\
\displaystyle{
4F^{\rm ext} =
-t_0^2 +2 t_0 \p_{t_0}F^{\rm ext} +
\sum_{k>0}(2-k)(t_k \p_{t_k}F^{\rm ext}
+\bar t_k \p_{\bar t_k}F^{\rm ext} )}.
\end{array}
\eeq
These formulas reflect the scaling
of moments as $z\to \lambda z$ with real $\lambda$:
$t_{k}^{\rm ext} \rightarrow \lambda^{2+k} t_{k}^{\rm ext}$,
$t_{k}^{\rm int} \rightarrow \lambda^{2-k} t_{k}^{\rm int}$.
The logarithmic moment $v_0$, under the same rescaling,
exhibits a more complicated behaviour:
$v_0 \rightarrow \lambda^2 v_0 +t_0 \lambda^2 \log \lambda^2$.
To get rid of the ``anomaly term'' $t_0^2$ one may modify
the tau-function by subtracting $\frac{1}{4}t_{0}^{2}\log t_{0}^{2}$.

\section{Dirichlet problem on the plane with a gap
\label{ss:gap}}

Consider the case when
the domain $D_{\rm int}$ shrinks to a segment of
the real axis. Then the interior Dirichlet
problem does not seem to make
sense anymore but the exterior one is still well-posed:  find a bounded
harmonic function in the complex plane such that it equals a given
function on the segment.
The problem admits an explicit solution
(see e.g. \cite{gap}).
Possible variations of the data are variation of the function on the
segment and the  endpoints of the
segment. Solution of this problem as well as its integrable structure may
be obtained from the formulas for a smooth domain as a result of
a singular limit when a smooth domain shrinks to the segment.

The  tau-function, obtained in this way,
is a partition function of the Hermitean one-matrix model
for the one-cut solution in the planar large $N$
limit~\cite{BIPZ}.

\subsection{Shrinking the domain: the limiting procedure}

Let us consider a family of curves $\gamma (\varepsilon)$
obtained from a given curve $\gamma$ by rescaling of the
$y$-axis as $y\to \varepsilon y$.
(If the curve $\gamma$ is given by an equation $P(x,y)=0$, then
$\gamma (\varepsilon)$ is given by $P(x,y/\varepsilon)=0$.)
We are interested in the limit $\varepsilon \to 0$,
$\gamma(0)$ being a segment of the real axis. We denote the
endpoints by $\alpha$, $\beta$.
Let $\Delta y(x)$ be the thickness
of the domain bounded by the curve
$\gamma (\varepsilon )$ at the point $x$
(see Fig.~\ref{fi:sosiska}).

\begin{figure}[tp]
\epsfysize=7cm
\centerline{\epsfbox{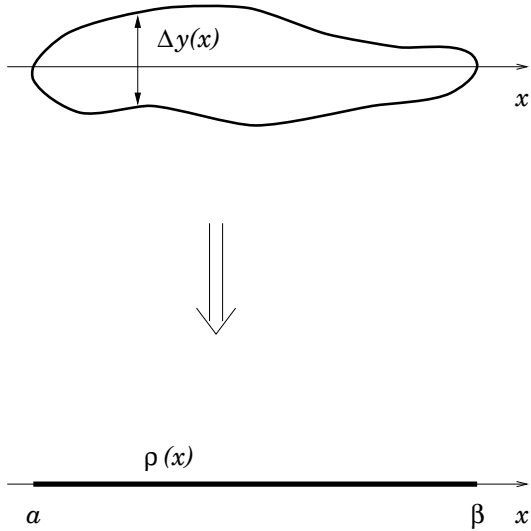}}
\caption{\sl A thin domain
stretched along the real axis with thickness $\Delta y(x)$
shrinks into a segment with density $\rho(x)$.}
\label{fi:sosiska}
\end{figure}

We introduce the function
$$
\rho(x)=\lim_{\varepsilon \to 0}
\frac{\Delta y(x)}{\varepsilon}.
$$
It is easy to see that in case of general position
this function can be represented as
\beq\label{gap1}
\rho (x)=\sqrt{(x-\alpha )(\beta -x)}\, M(x),
\eeq
where $M(x)$ is a smooth function regular at the edges.
So, instead of the space of contours $\gamma$ we have
the space of real positive functions $\rho (x)$
of the form (\ref{gap1}) with
a finite support $[\alpha , \beta ]$ (the endpoints
of which are not fixed but may vary).

Let us
use the first equation in
(\ref{expan12}) to define times as coefficients of the
expansion of the potential
$\Phi (x)=\Phi^{\rm ext}(x)$ generated by the thin domain with a
uniform charge density (and with a
point-like charge at the origin).  In
the leading order in $\varepsilon$ the
rescaled potential is
\beq
\label{phi1}
\phi(x)\equiv\lim_{\varepsilon
\to 0} \frac{\Phi (x)}{\varepsilon}=-\,
\frac{1}{\pi}\int_{\alpha}^{\beta}dx' \rho(x')\log \Bigl (
\frac{1}{x}-\frac{1}{x'}\Bigr )^2 = T_0 \log x^2 +\sum_{k\geq 1}T_k x^k.
\eeq
Comparing this with (\ref{expan12}), we get
\beq\label{Tktk}
\begin{array}{l}
T_0=\displaystyle{\lim_{\varepsilon\to 0}} \,\varepsilon^{-1}
t_{0} \,,
\\ \\
T_k=\displaystyle{\lim_{\varepsilon\to 0}} \,\varepsilon^{-1}
(t_{k}^{\rm ext} +\bar t_{k}^{\rm ext}) \,,
\;\;\;\;\;k\geq 1, \; k\neq 2,
\\ \\
T_2=\displaystyle{\lim_{\varepsilon\to 0}} \,\varepsilon^{-1}
(t_{2}^{\rm ext} +\bar t_{2}^{\rm ext} -1).
\end{array}
\eeq
Note that $T_{0}=\displaystyle{\frac{1}{\pi}
\int_{\alpha}^{\beta} \rho(x)dx}$ but
similar integral representations
for other times, $T_{k}=\displaystyle{\frac{2}{\pi k}
\int_{\alpha}^{\beta}\rho(x) x^{-k} dx}$, which formally
follow from (\ref{phi1}), are ill-defined. On the other hand,
harmonic moments of the interior
behave, in the scaling limit, as
$t_{k}^{\rm int}=\varepsilon k^{-1}\mu_k +O(\varepsilon^2)$,
where $\mu_k$ are
well-defined moments of the function $\rho$
on the segment:
\beq\label{gap5}
\mu_k =\frac{1}{\pi}
\int_{\alpha}^{\beta}\rho(x)x^k dx\,.
\eeq
Using integral formula (\ref{first}),
it is now straightforward to find the
scaling limit of the tau-function
$F^{\rm ext}$ for the exterior problem. Taking into account
(\ref{Tktk}), we introduce the function $F^{\rm cut}$
as follows:
\beq\label{Fcut}
\begin{array}{ll}
&F^{\rm ext}(\varepsilon t_0;\,
\varepsilon t_1,\,
\frac{1}{2}\!+\! \varepsilon t_2,\,
\varepsilon t_3, \,
\ldots \,;\,
\varepsilon \bar t_1, \,
\frac{1}{2}\!+\! \varepsilon \bar t_2,\,
\varepsilon \bar t_3, \,
\ldots \,)\, =
\\&\\
=& \varepsilon^2 \,
F^{\rm cut}(T_0 ; \,T_1, \,T_2, \, T_3 , \ldots )\, + O(\varepsilon^3),
\end{array}
\eeq
where $T_0 =t_0$, $T_{k}=t_k +\bar t_k$.
Since the second order derivatives of $F^{\rm ext}$ are
invariant under the rescaling (\ref{Fcut}),
the function $F^{\rm cut}$ obeys the Hirota
equations (\ref{Hir1})--(\ref{Hir4}) (as $F^{\rm ext}$ does)
where one has to
set $\p_{t_k}=\p_{\bar t_k}=\p_{T_k}$. The latter
means that $F^{\rm cut}$ is a solution of the
reduced dToda hierarchy (see e.g.\,\cite{KriW}).
This sort of reduction is
usually referred to as dispersionless Toda chain.
We conjecture that other types of reduction correspond,
in the same way, to shrinking of $D_{\rm int}$ to slit
domains of a more complicated form.

An integral representation for the $F^{\rm cut}$
(obtained as a limit of (\ref{first}))
reads:
$$
F^{\rm cut}=-\, \frac{1}{\pi^2}
\int_{\alpha}^{\beta}\!
\int_{\alpha}^{\beta}
\rho(x_1)\rho(x_2)
\log |x_{1}^{-1}-x_{2}^{-1}|
dx_1dx_2
$$
Represent this as
$F^{\rm cut}=\displaystyle{\frac{1}{2\pi}\int_{\alpha}^{\beta}
\rho(x)\phi(x)dx =\frac{1}{2}\sum_{k\geq 0}\mu_k T_k}$,
where $\mu_k$ are defined in (\ref{gap5}) and
$$
\mu_{0}=\frac{1}{\pi}
\int_{\alpha}^{\beta}\rho(x)\log x^2 dx\,.
$$
It is clear from the limit of (\ref{first11}) that
$\mu_k =\p_{T_k} F^{\rm cut}$ for $k\geq 0$.
Therefore, we obtain the relation
\beq\label{gap6}
2F^{\rm cut}=\sum_{k\geq 0}T_k \p_{T_k}F^{\rm cut},
\eeq
which means that $F^{\rm cut}$ is a homogeneous function
of degree 2.
This formula also means
that the tau-function $F^{\rm cut}$ is ``self-dual''
under the Legendre
transform with respect to $T_0 , T_1, T_2 , \ldots $.
However, the analog of the Legendre transform (\ref{Legendre})
we discussed in sect.~\ref{ss:Legendre}
does not include $T_0$, so the analog of the function
$F^{\rm int}$ is the function
$$
\begin{array}{lll}
E^{\rm cut}&=&
\displaystyle{\sum_{k\geq 1}T_k \mu_k -F^{\rm cut}}\, = \,
\displaystyle{
F^{\rm cut}-T_0 \frac{\p F^{\rm cut}}{\p T_0}}\, =
\\&&\\
&=&
\displaystyle{
-\, \frac{1}{\pi^2}
\int_{\alpha}^{\beta}\!
\int_{\alpha}^{\beta}
\rho(x_1)\rho(x_2)
\log |x_{1}-x_{2}|
dx_1dx_2},
\end{array}
$$
which is the electrostatic energy of the segment with the
charge density $\rho$ on it,
regarded as a function of the variables $T_0 , \mu_1 ,\mu_2 ,
\ldots$.
Properties of this function and its possible relation
to the Hamburger 1D moment problem are to be
further investigated.

\subsection{Conformal maps}

The conformal map
$f:{\bf C}\setminus [\alpha , \beta ]
\longrightarrow E$ from the plane with the gap onto the
exterior $E$ of the unit disk is given by the
explicit formula
$$
f(z)=\frac{2z -\alpha -\beta +2
\sqrt{(z-\alpha )(z-\beta )}}{\beta -\alpha }.
$$
All formulas which connect conformal maps and tau-function
remain true in this case, too. In particular,
expanding $f(z)$ as $z\to \infty$,
$f(z)=\displaystyle{\frac{4z}{\beta -\alpha}-
\frac{2(\alpha +\beta )}{\beta -\alpha}+O(z^{-1})}$,
we read from (\ref{1qa}) formulas for the endpoints of the cut:
$$
\beta -\alpha =4\exp \displaystyle{
\Bigl ( \frac{1}{2}\p^2_{T_0}F^{\rm cut}\Bigr )}\,,
\;\;\;\;\;
\beta +\alpha =2 \p^{2}_{T_0T_1}F^{\rm cut}.
$$
The Green function is expressed through the $f(z)$
by the same formula (\ref{G2}).

Set $w=f(z)$, then the inverse map is
$$
z(w)=\frac{\beta -\alpha}{4}(w+w^{-1})+\frac{\beta +\alpha}{2}.
$$
Clearly, $z(w)=\bar z(w^{-1})$, so the constraint on the
Lax functions is now $L(w)=\tilde L(w)$ which signifies
the reduction to the dispersionless Toda chain.

\subsection{Relation to matrix models}

In analogy with variations of the data of
the Dirichlet problem discussed in Sec.\,3, we
can consider the following problem: given a set of $T_k$,
$k\geq 0$, to find endpoints $\alpha , \beta$ of
the segment and the
density $\rho (x)$ as functions of these parameters.
This problem has ben appeared in studies of  the planar $N\to \infty$
limit of the Hermitean matrix model. In this case the function  $\rho (x)$
is a density of eigenvalues
\cite{BIPZ}. For completeness, we recall the
basic points.

Taking derivative of (\ref{phi1}), we get the equation
which connects the set of $T_k$ with $\alpha , \beta$ and
$\rho (x)$:
$$
\frac{2}{\pi}\,\mbox{v.p.} \! \int_{\alpha}^{\beta}
\frac{\rho (x') dx'}{x'-x} =V'(x)\,,
$$
where
\be
\label{V}
V(x)=\phi (x)-T_0 \log x^2 =\sum_{k\geq 1}T_k x^k\,.
\ee
Consider the function
$$
W(z)=\frac{1}{\pi}\int_{\alpha}^{\beta}
\frac{\rho (x)dx}{z-x}\,.
$$
It is analytic in the complex plane with the cut
$[\alpha , \beta ]$. As $z\to \infty$, it behaves as
$W(z)=T_0 z^{-1}+O(z^{-2})$. On the cut, its boundary values
are $W(x\pm i0)=-\frac{1}{2}V'(x)\mp i\rho (x)$. The function
$W(z)$ is uniquely defined by these analytic properties.
So, to find $\rho$ one should find the holomorphic function
from its mean value on the cut. The result is given
by the explicit formula
\beq\label{gap4}
W(z)=\frac{1}{4\pi i}\oint \frac{d\zeta V'(\zeta )}{\zeta -z}
\,\frac{\sqrt{(z-\alpha )(z-\beta )}}{\sqrt{
(\zeta -\alpha )(\zeta -\beta )}}\,,
\eeq
where the contour encircles the cut but not the
point $z$. The endpoints of the cut are fixed by
comparing the leading terms of (\ref{gap4}) with the
required asymptote of the $W(z)$ as $z\to \infty$.
This leads to the hodograph-like formulas
$$
\frac{1}{2\pi i}\oint \frac{V'(z)dz}{\sqrt{
(z-\alpha )(z-\beta )}}=0,
\;\;\;\;
\frac{1}{2\pi i}\oint \frac{zV'(z)dz}{\sqrt{
(z-\alpha )(z-\beta )}}=-2T_0,
$$
which implicitly determine $\alpha$, $\beta$ as functions
of $T_k$.

It follows from the above that $F^{\rm cut}$
coincides with the one-cut
free energy of the Hermitean one-matrix model
with potential (\ref{V})
in the planar large $N$ limit.
(As a matter of fact,
it was shown in \cite{GMMMO} that
the partition function of the matrix model at finite $N$ is
the tau-function of the Toda chain hierarchy with dispersion.)
Combining (\ref{gap6}) with the $\varepsilon \to 0$
limit of the second
relation in (\ref{hom1}), we obtain
$$
\sum_{k\geq 1}kT_{k} \p_{T_k}F^{\rm cut}+T_0^2 =0.
$$
This identity is known as the Virasoro $L_0$-constraint on the
dispersionless tau-function $F^{\rm cut}$ \cite{GMMMO,KriW}.
Other Virasoro constraints can be
obtained in the $\varepsilon \to 0$ limit from the
$W_{1+\infty}$-constraints on the tau-function $F^{\rm ext}$.
We do not discuss them here.

\subsection{An example: Gaussian matrix model}

If only first three
variables are nonzero (i.e., $T_0, T_1, T_2 \neq 0$)
function $F^{\rm cut}$ can be found explicitly.
Consider a family of ellipses with axes $l$ and
$\varepsilon s$
centered at some point $x_0$ on real axis:
$$
\frac{(x-x_0)^2}{l^2}+\frac{y^2}{\varepsilon ^2 s^2}
=\frac{1}{4}.
$$
The harmonic moments, as $\varepsilon \to 0$, are
(see Appendix in ref.\,\cite{W-Z}):
$t_0 =\frac{1}{4}\varepsilon sl$,
$t^{\rm ext}_{1} =2\varepsilon x_0 s l^{-1} +O(\varepsilon^2)$,
$t^{\rm ext}_{2} =\frac{1}{2}
-\varepsilon sl^{-1} +O(\varepsilon^2)$ and all other moments
of the complement to ellipse vanish.
 From (\ref{Tktk}) we read values of the rescaled variables:
$$
T_0 =\frac{1}{4}(\beta -\alpha )s\,,
\;\;\;\;
T_1 =2\, \frac{\beta +\alpha}{\beta -\alpha}\,s\,,
\;\;\;\;
T_2 =-\,\frac{2}{\beta -\alpha}\, s\,,
$$
and all the rest are zero. Here we have expressed
times through the endpoints $\alpha =x_0 -\frac{1}{2}l$,
$\beta =x_0 +\frac{1}{2}l$ of the segment
and the extra parameter $s$.
The density function is
$$
\rho (x)=-T_2 \sqrt{(\beta -x)(x-\alpha )}.
$$
(Note that $T_2 <0$.)

Using explicit form of the tau-funtion for ellipse
\cite{W-Z,K-K-MW-W-Z},
$$
\left. \phantom{F^a} F^{\rm ext}\right |_{\rm ellipse}
=\frac{1}{2}t_{0}^{2}\log
\frac{t_{0}}{1-4t_2 \bar t_2}
-\frac{3}{4}t_{0}^{2} +t_0 \,
\frac{t_{1}\bar t_1 +t_{1}^{2}\bar t_2
+\bar t_{1}^{2}t_2}{1-4t_2 \bar t_2}\,,
$$
and the scaling procedure (\ref{Fcut}), we find
$$
F^{\rm cut}(T_0 , T_1 , T_2 , 0,0,\ldots )=
\frac{1}{2}T_{0}^{2}\log \left (
\frac{T_0}{-2T_2}\right )
-\frac{3}{4}T_{0}^{2}
-\frac{T_0 T_{1}^{2}}{4T_2}.
$$
This is expression for the free energy
of the Gaussian matrix model in the planar
large $N$ limit \cite{BIPZ}.
The same result can be obtained from the
integral formula for $F^{\rm cut}$.

\section*{Acknowledgments}
We acknowledge useful dis\-cus\-sions with A.Boyarsky,
L.Chekhov, B.Dub\-ro\-vin, P.Di Fran\-ces\-co
M.Mi\-ne\-ev-\-Wein\-stein,
V.Ka\-za\-kov, A.A.Ki\-ril\-lov,
I.Kos\-tov, A.Po\-lya\-kov, O.Ru\-chay\-skiy,
and especially
with I.Kri\-che\-ver and L.Takh\-ta\-jan. The work of A.M. and A.Z. was
partially supported by CRDF grant RP1-2102
and RFBR grant No.~00-02-16477,
P.W. and A.Z. have been supported by grants NSF DMR
9971332 and MRSEC NSF DMR 9808595.
A.M. was partially supported by
INTAS grant 97-0103
and grant for support of scientific schools
No.~00-15-96566.
A.Z. was partially supported
by grant INTAS-99-0590, by
RFBR grant 98-01-00344 and grant
for support of scientific schools
No.~00-15-96557.


\begin{thebibliography}{66}


\bibitem{C-H} A.~Hurwitz and R.~Courant,
{\it Vorlesungen \"uber allgemeine Funktionentheorie und
elliptische Funktionen. Herausgegeben und erg\"anzt durch
einen Abschnitt \"uber geometrische Funktionentheorie},
Springer-Verlag, 1964 (Russian translation, adapted
by M.A.~Evgrafov: {\it Theory of functions}, Nauka, Moscow,
1968).


\bibitem{M-W-Z}
M.~Mineev-Weinstein, P.B.~Wiegmann and A.~Zabrodin,
Phys. Rev. Lett. {\bf 84} (2000) 5106,
e-print archive: nlin.SI/0001007.

\bibitem{W-Z}
P.B.~Wiegmann and A.~Zabrodin, Commun. Math. Phys.
{\bf 213} (2000) 523, e-print archive: hep-th/9909147.


\bibitem{gravity}
A.~Hanany, Y.~Oz and R.~Plesser,
Nucl. Phys. {\bf B425} (1994) 150-172; \\
K.~Takasaki,
Commun. Math. Phys. {\bf 170} (1995) 101-116; \\
T.~Eguchi and H.~Kanno, Phys. Lett. {\bf 331B} (1994) 330.

\bibitem{2matrix}
J.M.Daul, V.A.Kazakov and I.K.Kostov,  Nucl. Phys.
{\bf B409} (1993) 311-338; \\
L.Bonora and C.S.Xiong,
Phys. Lett. {\bf B347} (1995) 41-48.


\bibitem{GT1}J.~Gibbons and S.~P.~Tsarev,
Phys. Lett. {\bf 211A} (1996)
19-24; ibid {\bf 258A} (1999) 263.

\bibitem{Hadamard}
J.~Hadamard, {\it M\'em. pr\'esent\'es par
divers savants \`a l'Acad. sci.},
{\bf 33} (1908).

\bibitem{K-K-MW-W-Z} I.K.~Kostov, I.M.~Krichever, M.~Mineev-Weinstein,
P.B.~Wiegmann and A.~Zabrodin, \emph{$\tau$-function for analytic
curves},
Random matrices and their applications, MSRI publications,vol.40,
Cambridge Academic Press, 2001, e-print archive: hep-th/0005259.




\bibitem{Kriunp}
I.~Krichever, unpublished.

\bibitem{T} L.~Takhtajan, {\it Free bosons and tau-functions
for compact Riemann surfaces and closed smooth Jordan curves.
I. Current correlation functions},
e-print archive: math.QA/0102164.



\bibitem{Hille}
E.~Hille, {\it Analytic function theory},
v.II, Ginn and Company, 1962.


\bibitem{KriW}I.M.~Krichever,
Funct. Anal Appl. {\bf 22} (1989) 200-213;\\
Commun. Pure. Appl. Math. {\bf 47} (1992) 437,
e-print archive: hep-th/9205110.


\bibitem{Takashi} T.~Takebe, Adv. Series in Math. Phys. {\bf 16}
(1992), Proceedings of RIMS Research Project 1991, 923-940.

\bibitem{Shiota} T.~Shiota, Invent. Math. {\bf 83} (1986) 333-382.

\bibitem{Douglas} M.~Douglas, Phys. Lett. {\bf B238} (1990) 176.

\bibitem{Crete} M.~Mineev-Weinstein and A.~Zabrodin,
Proceedings of the Workshop NEEDS 99 (Crete, Greece, June 1999),
e-print archive: solv-int/9912012.

\bibitem{KMMM} S.~Kharchev, A.~ Marshakov, A.~Mironov and A.~Morozov,
Mod. Phys. Lett. {\bf A8} (1993) 1047-1061, e-print archive:
hep-th/9208046.


\bibitem{Sato} M.~Sato, \emph{Soliton
Equations and Universal Grassmann
Manifold} Math.~Lect.~Notes Ser.,
Vol.~18, Sophia University, Tokyo (1984);\\
E. Date, M. Jimbo, M.Kashiwara and T. Miwa, {\it
Transformation groups for soliton equations}, in: Nonlinear
Integrable Systems, eds. M. Jimbo and T. Miwa, Singapore,
World Scientific, 1983.


\bibitem{C-K}
J. Gibbons and Y. Kodama, Proceedings of NATO ASI ``Singular
Limits of Dispersive Waves'', ed. N. Ercolani,
London -- New York, Plenum, 1994;\\
R.~Carroll and Y.~Kodama, J. Phys. A: Math. Gen.
{\bf A28} (1995) 6373.

\bibitem{T-T} K.~Takasaki and T.~Takebe,
Rev. Math. Phys. {\bf 7} (1995) 743-808.

\bibitem{WDVV}
R.~Dijkgraaf, E.~Verlinde and H.~Verlinde, Nucl. Phys. {\bf B352}
(1991) 59.

\bibitem{MMM}
A.~Marshakov, A.~Mironov and A.~Morozov,
Phys. Lett. {\bf B389} (1996) 43-52,
e-print archive: hep-th/9607109.

\bibitem{BMRWZ}
A.~Boyarsky, A.~Marshakov, O.~Ruchayskiy, P.~Wiegmann
and A.~Zabrodin, to be published in Phys. Lett. B,
e-print archive: hep-th/0105260.

\bibitem{gap} F.Gakhov, {\it Boundary problems},
Nauka, Moscow, 1977 (in Russian);\\
A.Bitsadze, {\it Foundations of the theory of analytic
functions of a complex variable}, Nauka, Moscow, 1984
(in Russian)

\bibitem{BIPZ} E.~Br\'ezin, C.~Itzykson, G.~Parisi and
J.-B.~Zuber, Commun. Math. Phys. {\bf 59} (1978) 35.

\bibitem{GMMMO} A.~Gerasimov, A.~Marshakov, A.~Mironov,
A.~Morozov and A.~Orlov, Nucl. Phys. {\bf B357} (1991)
565-618.

\end{thebibliography}
\end{document}